\documentclass[letter,scriptaddress,twocolumn, prl,showkeys]{revtex4}
\usepackage{graphicx}
\usepackage[caption=false]{subfig}
\usepackage{siunitx}

\begin{document}
\title{Observation of Stimulated Hawking Radiation in an Optical Analogue}
\author{Jonathan Drori$^1$}
\author{Yuval Rosenberg$^1$}
\author{David Bermudez$^2$}
\author{Yaron Silberberg$^1$}
\author{Ulf Leonhardt$^1$}
\affiliation{
\normalsize{
$^1$Weizmann Institute of Science, Rehovot 7610001, Israel\\
$^2$Departamento de F\'{i}sica, Cinvestav, A.P. 14-740, 07000 Ciudad de M\'{e}xico, Mexico}
}
\begin{abstract}
The theory of Hawking radiation can be tested in laboratory analogues of black holes. We use light pulses in nonlinear fiber optics to establish artificial event horizons. Each pulse generates a moving perturbation of the refractive index via the Kerr effect. Probe light perceives this as an event horizon when its group velocity, slowed down by the perturbation, matches the speed of the pulse. We have observed in our experiment that the probe stimulates Hawking radiation, which occurs in a regime of extreme nonlinear fiber optics where positive and negative frequencies mix.
\end{abstract}
\date{\today}

\maketitle
In 1974 Stephen Hawking published his best--known paper \cite{Hawking} where he theorized that black holes are not entirely black, but radiate due to the quantum nature of fields. Hawking's paper confirmed Jacob Bekenstein's idea \cite{Bekenstein} of black-hole thermodynamics that subsequently became a decisive test for theories of quantum gravity. Yet Hawking radiation has been a theoretical idea itself; its chances of observation in astrophysics are astronomically small indeed \cite{Oom}. 

In 1981 William Unruh suggested \cite{Unruh} an analogue of Hawking's effect \cite{Hawking} that, in principle, is observable in the laboratory. Unruh argued that a moving quantum fluid with nonuniform velocity --- liquid Helium \cite{Volovik} was the only choice at the time --- establishes the analogue of the event horizon when the fluid exceeds the speed of sound. This is because sound waves propagating against the current can escape subsonic flow, but are dragged along in spatial regions of supersonic flow. Mathematically, the moving fluid establishes a space--time metric that is equivalent to the geometry of event horizons \cite{Unruh, Volovik, Review}. So the analogue to quantum fields in space--time geometries should exhibit the equivalent of Hawking radiation as well, Hawking sound in Unruh's case \cite{Unruh}.

With this \cite{Unruh} and other \cite{Review} analogues one can investigate the influence of the extreme frequency shift at horizons, shifts beyond the Planck scale \cite{Jacobson}. The particles of Hawking radiation appear to originate from extreme frequency regions where the physics is unknown. In analogues of the event horizon, instead of the unknown physics beyond the Planck scale, the known frequency response of the materials involved regularize the extreme frequency shift \cite{Jacobson, Unruh2}. Analogues are thus a testing ground for the potential influence of trans-Planckian physics on the Hawking effect. 

In 2000 horizon analogues began to become the subject of serious experimental effort and to diversify into various areas of modern physics. While none of the first proposals \cite{LP, Garay} were directly feasible, they inspired experiments on horizons in optics \cite {Philbin, Faccio1, Faccio2, Genov, Rivka1, Rivka2}, ultra-cold quantum gases \cite{Stein1, Stein2, Stein3,Stein4}, polaritons \cite{Nguyen} and water waves \cite{Rousseaux, Jannes, Weinfurtner1, Euve1, Euve2, Weinfurtner2}. Yet despite admirable experimental progress, there is still no clear--cut demonstration of quantum Hawking radiation. The optical demonstration \cite{Faccio1} turned out to be the horizon--less emission from a superluminal refractive--index perturbation \cite{Petev}. The demonstration \cite{Stein2} of black hole lasing in Bose-Einstein condensates (BEC) was disputed \cite{BHLdebate,SteinReply} with overwhelming arguments \cite{BHLdebate}, and the demonstration of Hawking radiation in BECs \cite{Stein3} appears to suffer from similar problems \cite{Parola,LeoAnnalen}, with the possible exception of Ref.~\cite{Stein4}.

Here we report on clear measurements of the stimulated Hawking effect in optics. This is not a full demonstration of quantum Hawking radiation yet, but it already gives quantitative experimental results on the spontaneous Hawking effect. It represents the next milestone following the demonstration of frequency shifting at horizons \cite{Philbin} and negative resonant radiation \cite{Faccio2}. From an optics perspective, it establishes the regime of extreme nonlinear fiber optics where controlled conversions between positive and negative frequencies occur.

In optical analogues \cite{Philbin}, an intense ultrashort light pulse in a transparent medium creates a perturbation $\delta n$ of the refractive index due to the Kerr effect \cite{Agrawal} that travels with the pulse. In a co--moving frame the pulse stands still and increases the local refractive index $n$, reducing the velocity of itself and other light, while the material appears to be moving against the pulse. For probe light present, the pulse establishes horizons where its group velocity $u$ matches the group velocity $c/(n+\omega\, dn/d\omega)$ of the probe. A black--hole horizon is formed in the leading end of the pulse, and a white--hole horizon in the trailing end \cite{Book}. In the spontaneous Hawking effect, the probe consists of vacuum fluctuations of the electromagnetic field, while in the stimulated effect, the probe has a coherent amplitude.

\begin{figure}[b]
	\includegraphics[width=1\linewidth]{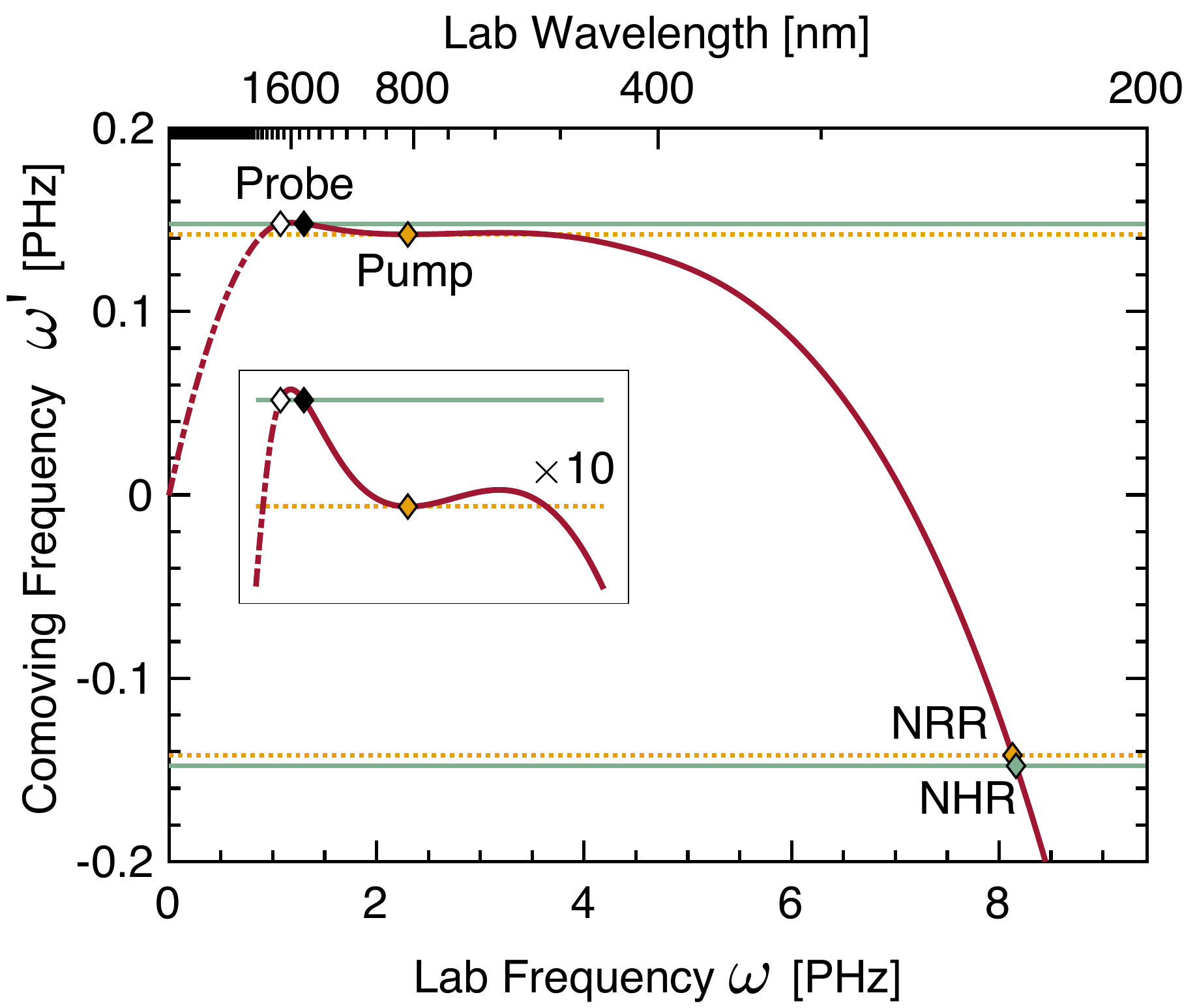}%
\caption{Doppler curve. Plot of $\omega'$ given by Eq.~(\ref{doppler}) for $n(\omega)$ and $u$ of our fiber (solid curve: $n$ determined from measurements \cite{Details}, dashed--dotted: $n$ extrapolated). The pump pulse sits at a local minimum and the horizon at a local maximum \cite{Faccio2}; $\omega'$ is conserved during pump--probe interaction (horizontal lines). The probe light (black and white diamonds) is incident with frequencies lower or higher than the horizon, experiencing the analogue of a black or a white hole. Both incident and outgoing Hawking partner has $-\omega'$ of the probe (lower line) intersecting the Doppler curve where we expect negative Hawking radiation (NHR, Fig.~\ref{results}c). The pump itself creates negative resonant radiation (NRR) at the intersection of its $-\omega'$ (lower dotted line) with the Doppler curve \cite{Faccio2}.}
\label{dopplercurve}
\end{figure}

Consider the probe in the co--moving frame. There the pulse is stationary and hence the co--moving frequency $\omega'$ of the probe is a conserved quantity. A pair of Hawking quanta thus consists of a photon with positive $\omega'$ and a partner with the exact opposite, $-\omega'$, such that the sum is zero. The time--dependent annihilation operators $\hat{b}_\pm$ of the outgoing radiation are given \cite{Book} by the Bogoliubov transformations of the time--dependent ingoing  $\hat{a}_\pm$ as
\begin{equation}
\hat{b}_\pm=\alpha\hat{a}_\pm+\beta\hat{a}_\mp^\dagger 
\label{bogo}
\end{equation}
with constant $\alpha,\beta$ and $|\alpha|^2-|\beta|^2=1$, where $\pm$ refers to the sign of $\omega'$. The ingoing field is incident in the material at rest, {\it i.e.}\ in the laboratory frame. This implies \cite{Book} that the $\hat{a}_\pm$ oscillate with positive laboratory frequencies $\omega$. To see how and for which positive $\omega$ negative co--moving frequencies $\omega'$ appear, consider the Doppler effect: 
\begin{equation}
\omega'=\gamma\left(1-n\frac{u}{c}\right)\omega 
\label{doppler}
\end{equation}
where $u$ denotes the group velocity of the pulse and $\gamma^{-2}=1-{u^2}/{c^2}$. For positive laboratory frequencies, $\omega'$ is positive when the phase velocity $c/n$ is faster than the pulse, which in our system is the case in the infrared (IR) (Fig.~\ref{dopplercurve}). The co--moving frequency $\omega'$ is negative when $u$ exceeds $c/n$, which occurs in the ultraviolet (UV) (Fig.~\ref{dopplercurve}).  Making measurements in these spectral regions gives us data on the Hawking effect. In particular, an IR probe with $\langle\hat{a}_-\rangle=0$ stimulates the UV signal $\langle\hat{b}_-\rangle=\beta\langle\hat{a}_+\rangle^*$ that proves the existence of the effect and gives the spontaneous photon number $|\beta|^2$ if the amplitude $\langle\hat{a}_+\rangle$ interacting with the pulse is known. 

Furthermore, while the ingoing modes oscillate with positive laboratory frequencies, the outgoing modes must contain negative--frequency contributions due to the Hermitian conjugation in the Bogoliubov transformation, Eq.~(\ref{bogo}). This combination of positive and negative frequencies in the laboratory frame differs from ordinary optical parametric amplification \cite{OPA} and is only possible in a regime of extreme nonlinear optics with few--cycle pulses beyond the slowly--varying envelope approximation \cite{Agrawal}. Only in this extreme regime $\beta$ is sufficiently large to be detectable. 

\begin{figure}[b]
	\includegraphics[width=1\linewidth]{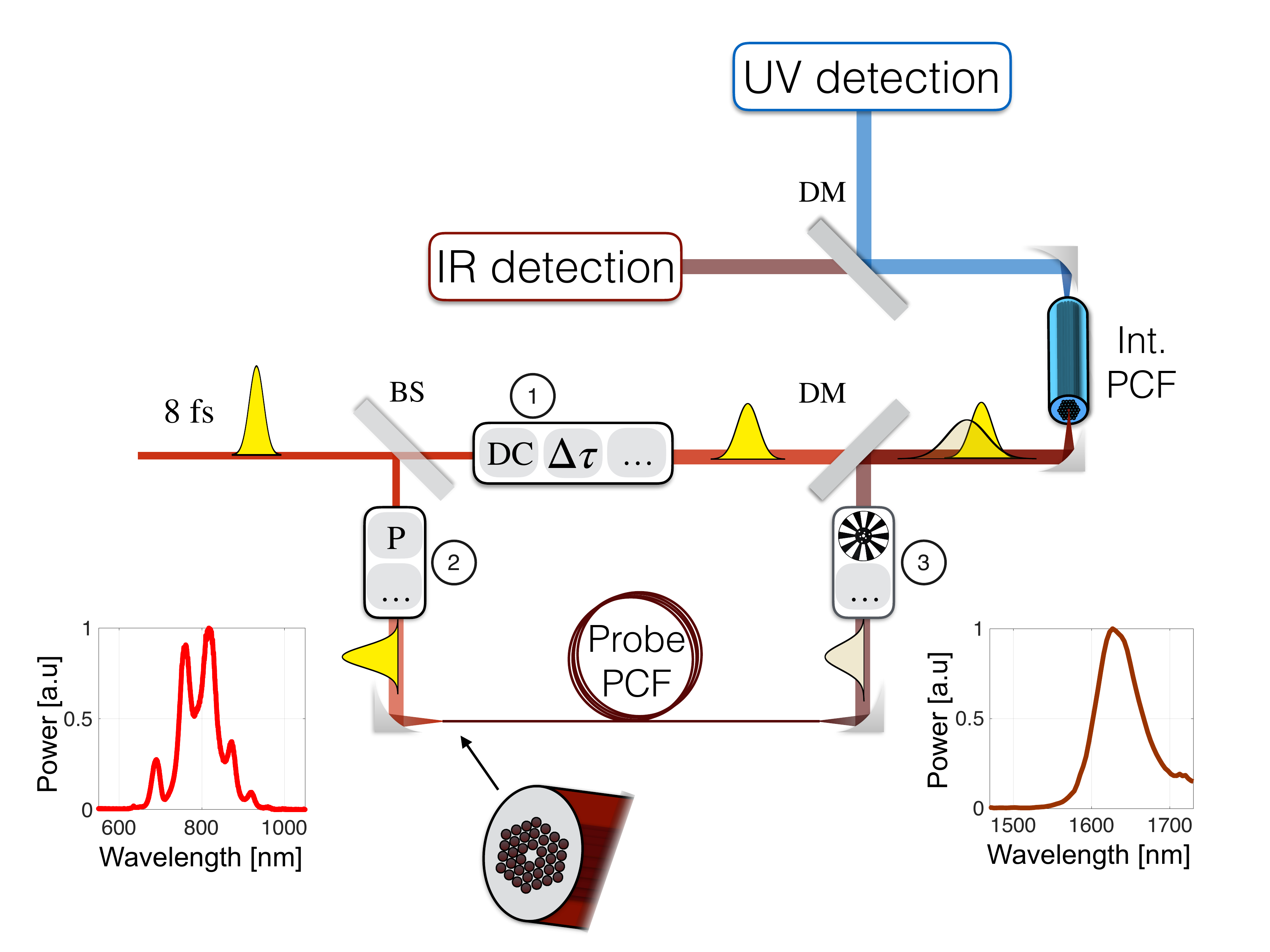}%
\caption{Experimental setup. The starting point (left) are light pulses of \SI{8}{\femto\second} duration at \SI{800}{\nano\meter} carrier wavelength. The inset shows the pulse spectrum. At the 50:50 beam splitter (BS) each pulse is distributed to two channels. In \textcircled{\raisebox{-0.9pt}{1}} the pulse is dispersion--compensated and delayed before being combined with the probe pulse that is prepared in the other arm. The intensity of the other split pulse is tuned \textcircled{\raisebox{-0.9pt}{2}} by a half--wave plate and a polarizer. It is coupled into the Probe Photonic Crystal Fiber (PCF) by a parabolic mirror to be Raman--shifted in its wavelength depending on the initial intensity. The right inset shows a typical spectrum after Raman shifting. In \textcircled{\raisebox{-0.9pt}{3}} the train of probe pulses is modulated by a chopper wheel before being combined with the pump pulse at a dichroic mirror (DM). Pump and probe enter the Interaction PCF via a parabolic mirror. The resulting light is collimated and distributed (DM) to the IR and UV detection stations.}
\label{scheme}
\end{figure}

Figure~\ref{scheme} shows our experimental setup. We perform a pump--probe experiment: the pulse creating the moving refractive--index perturbation is called pump, and its effect is probed by a probe pulse we derive from the same source as the pump. The pump pulses are of \SI{8}{\femto\second} duration at \SI{800}{\nano\meter} free-space carrier wavelength (produced by a Thorlabs Octavius oscillator). They are coupled into a \SI{7}{mm} photonic-crystal fiber (PCF) (NKT NL-1.5-590). In this fiber, probe pulses of $\approx$ \SI{50}{\femto\second} duration and tuneable carrier wavelength may interact with the pump. The probe pulses have been generated by Raman shifting \cite{Agrawal} in  a \SI{1}{\meter}  PCF (NKT NL-1.7-765) after reflection off the original master pulses by a 50:50 beamsplitter. We take advantage of the intensity dependence of the Raman effect to tune them over the wavelength range from \SI{800}{\nano\meter} to \SI{1620}{\nano\meter} by small intensity changes. The output of the pump--probe interaction is distributed via a dichroic mirror and spectral and spatial filters to two detection stations, a commercial spectrometer (Avantes AvaSpec-NIR256-1.7) for the IR and, for the UV, a prism--based tuneable monochromator and a photomultiplier tube (Hamamatsu H8259) as detector.

\begin{figure}[t]
	\centering\subfloat{\includegraphics[width=1\linewidth]{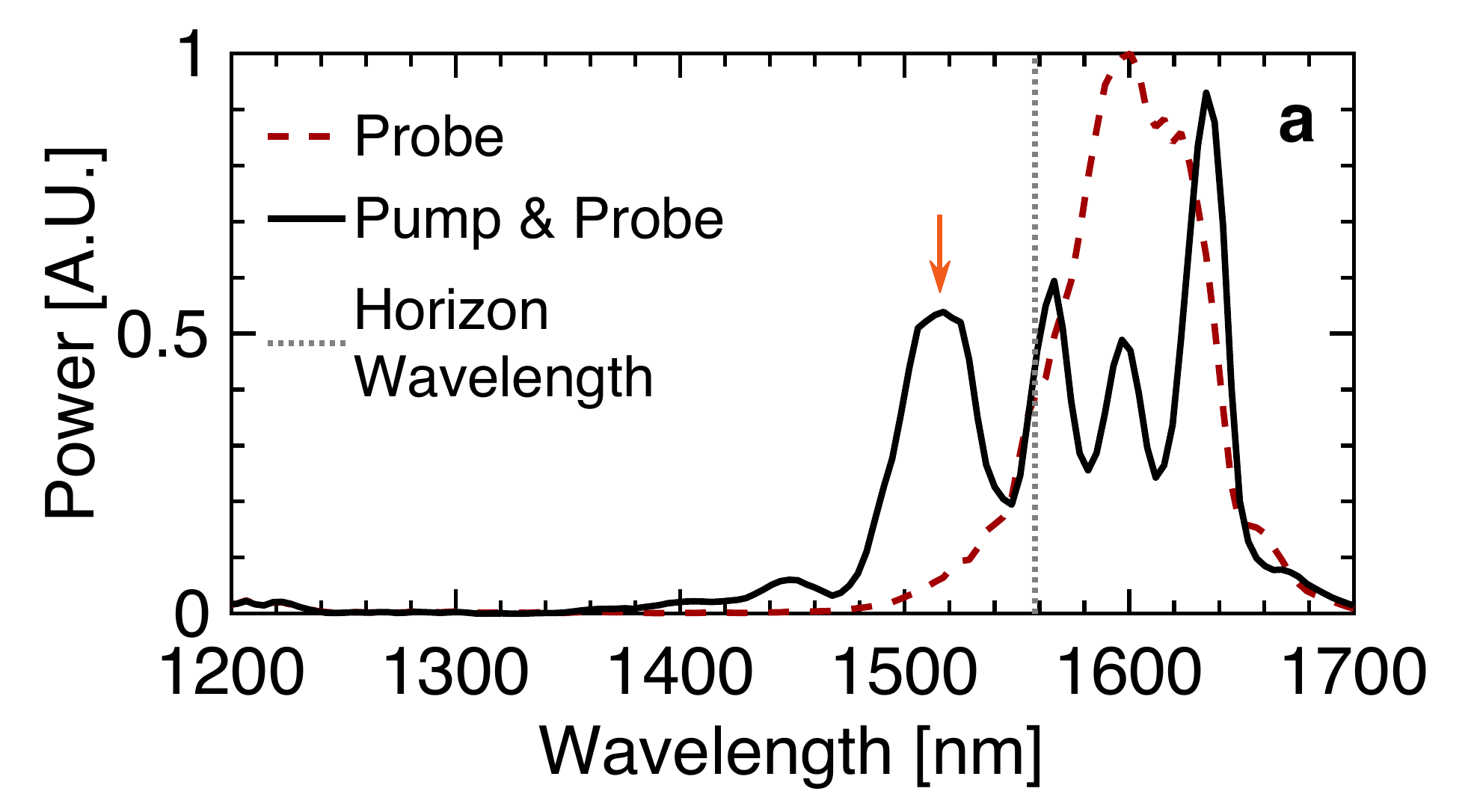} \label{results a}}\\
	\centering\subfloat{\includegraphics[width=1\linewidth]{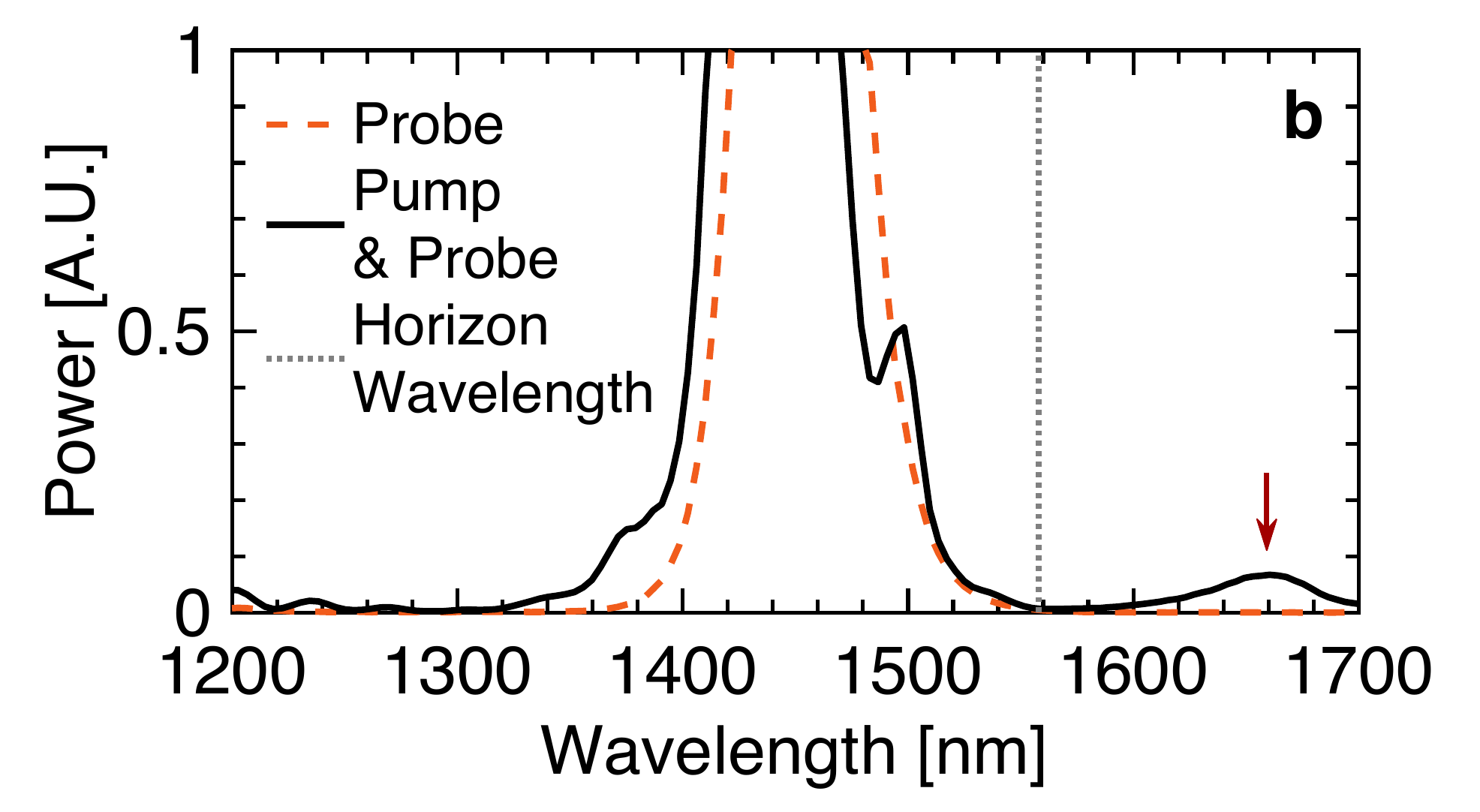}\label{results b}}\\
	\centering\hspace{-0.0cm}\subfloat{\includegraphics[width=1\linewidth]{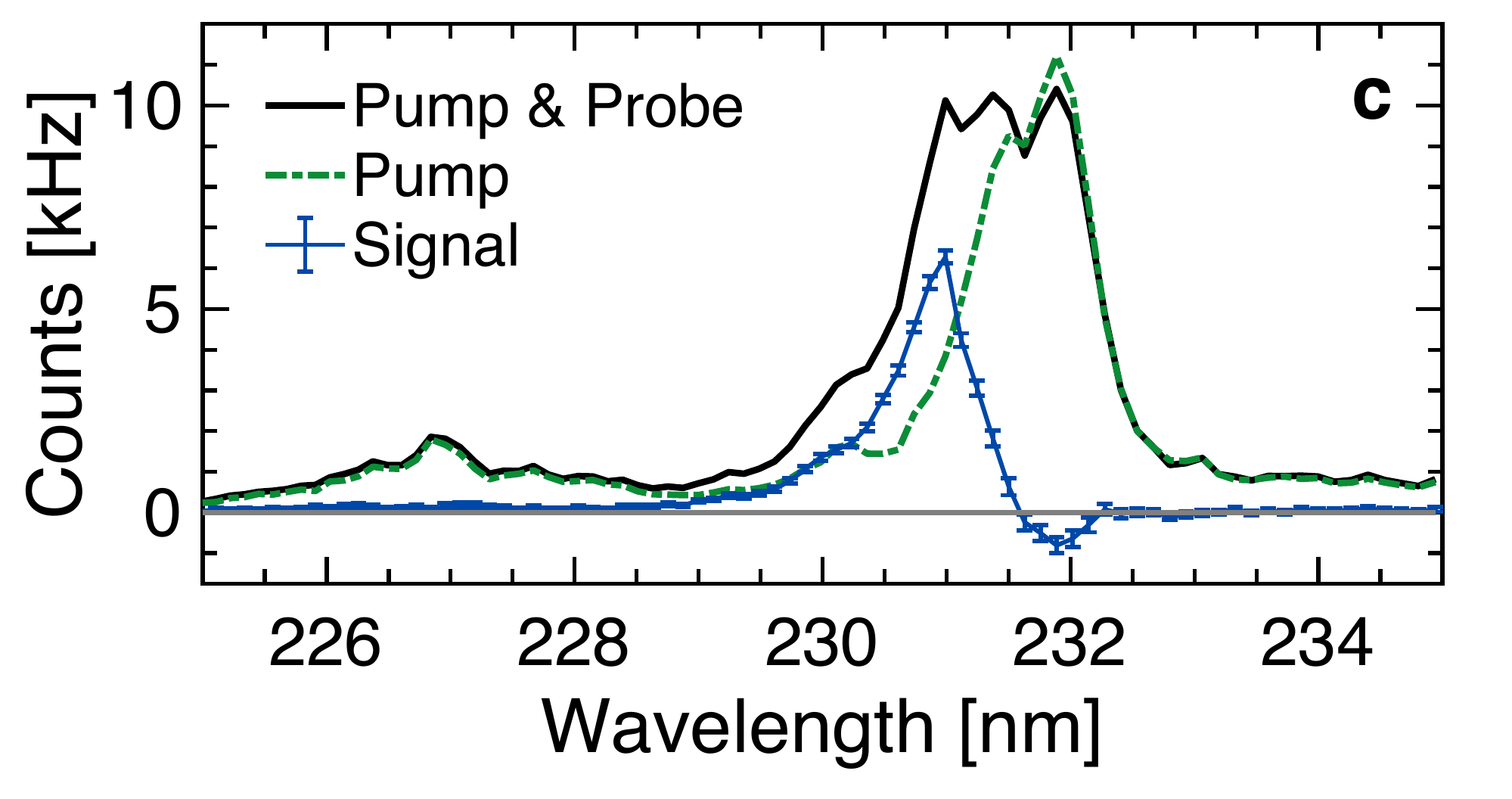}\label{results c}}
\caption{Experimental results. {\bf a}: Spectrum of the IR probe (solid curve) after the interaction with the pump for an initial probe (dashed curve) tuned to the red side of the horizon wavelength (dotted line). The probe has been blue--shifted (arrow) and also spectrally modulated. {\bf b}: Spectrum (solid curve) after interaction for an initial probe (dashed line) tuned to the blue side of the horizon (dotted line). The figure shows a distinct red shift (arrow). {\bf c}: UV spectrum for the \SI{1450}{\nano\meter} probe shown in b interacting with the pump (solid curve) and for the pump alone (dashed and dotted). The difference produces a clear signal (curve with $2\sigma$ error bars) we interpret as stimulated negative Hawking radiation (Fig.~\ref{dopplercurve}, NHR).}
\label{results}
\end{figure}

Some representative detection results are shown in Fig.~\ref{results}. To understand them we note the following. For the probe the group--velocity dispersion is normal --- the group velocity increases with increasing wavelength. At the horizon wavelength the group velocity of the probe matches the pulse velocity, so the probe is faster than the pump for longer wavelengths and slower for shorter wavelengths. Therefore, when the probe is tuned to the red side of the horizon wavelength (Fig.~\ref{results a}) it runs into the white--hole horizon and is blue--shifted \cite{Philbin,Note} (Fig.~\ref{results a}). The probe on the blue side (Fig.~\ref{results  b}) is slower then the pulse, experiences a black--hole horizon and is red-shifted \cite{Philbin,Note} (Fig.~\ref{results  b}). The red--shifting produces a clearer signal than the blue--shifting, although of lower magnitude, because, due to the Raman effect \cite{Agrawal}, the pump pulse de--accelerates \cite{Agrawal} such that the blue--shifted probe light interacts longer with the white--hole horizon, and with more complicated dynamics (producing the spectral modulations of Fig.~\ref{results}a). In extreme cases, the probe may even get trapped by the pulse \cite{GS}. 

Figure~\ref{results c} shows results of the UV detection with and without the probe. With the probe off, one sees a clear peak at \SI{231.9}{\nano\meter} wavelength that corresponds to the negative resonant radiation (NRR, Fig.~\ref{dopplercurve}) stimulated by the pump itself \cite{Faccio2}. With the probe on, the peak gets visibly broader. Taking the difference reveals an additional signal peaked at \SI{231}{\nano\meter} (for a probe wavelength of \SI{1450}{\nano\meter}). This feature, stimulated by the probe, we believe is the negative--frequency component of stimulated Hawking radiation.

\begin{figure}[t]
	\centering\hspace{-0.15cm}\subfloat{\includegraphics[width=1\linewidth]{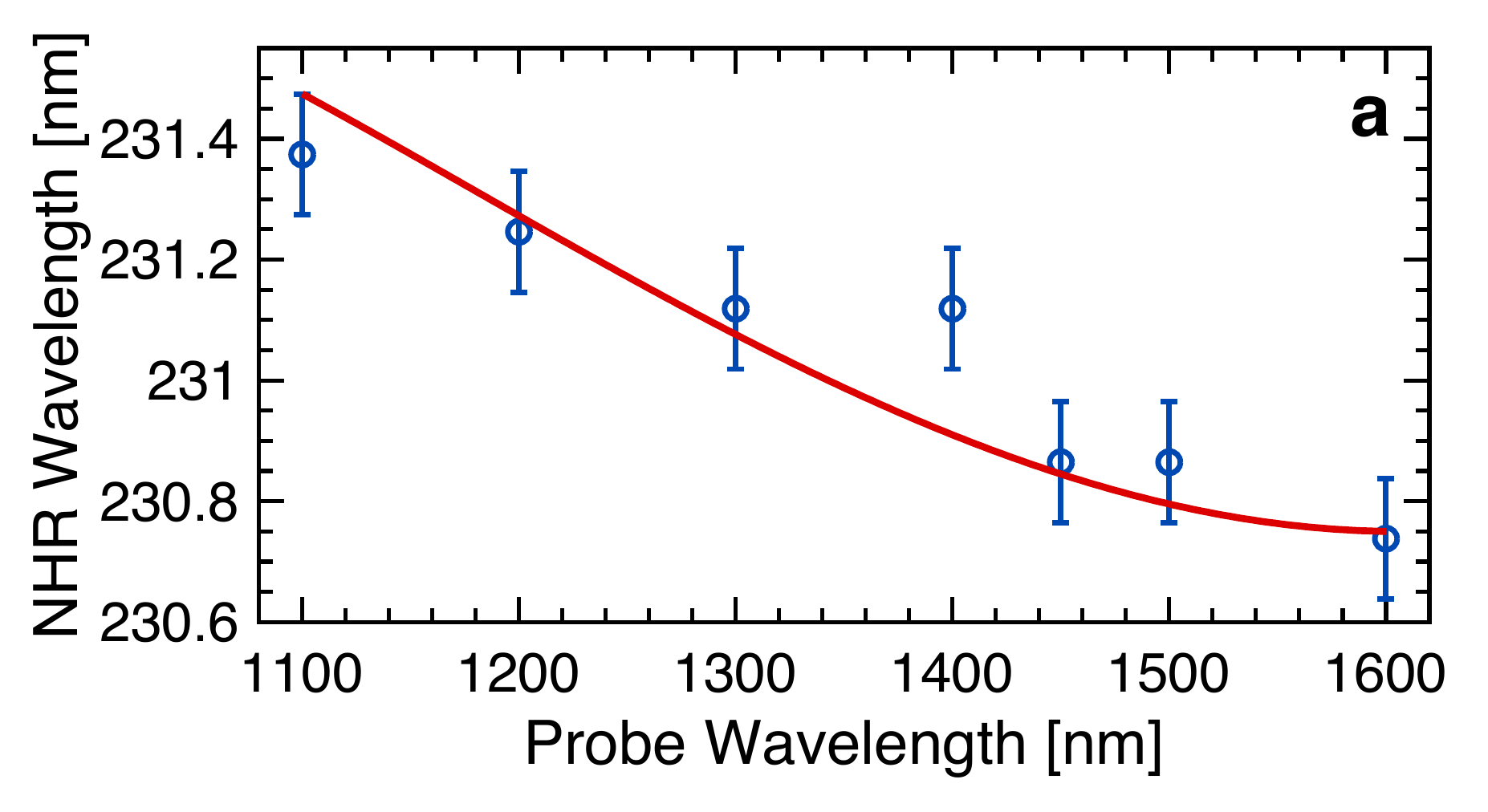} \label{test a}}\\
	\centering\subfloat{\includegraphics[width=1\linewidth]{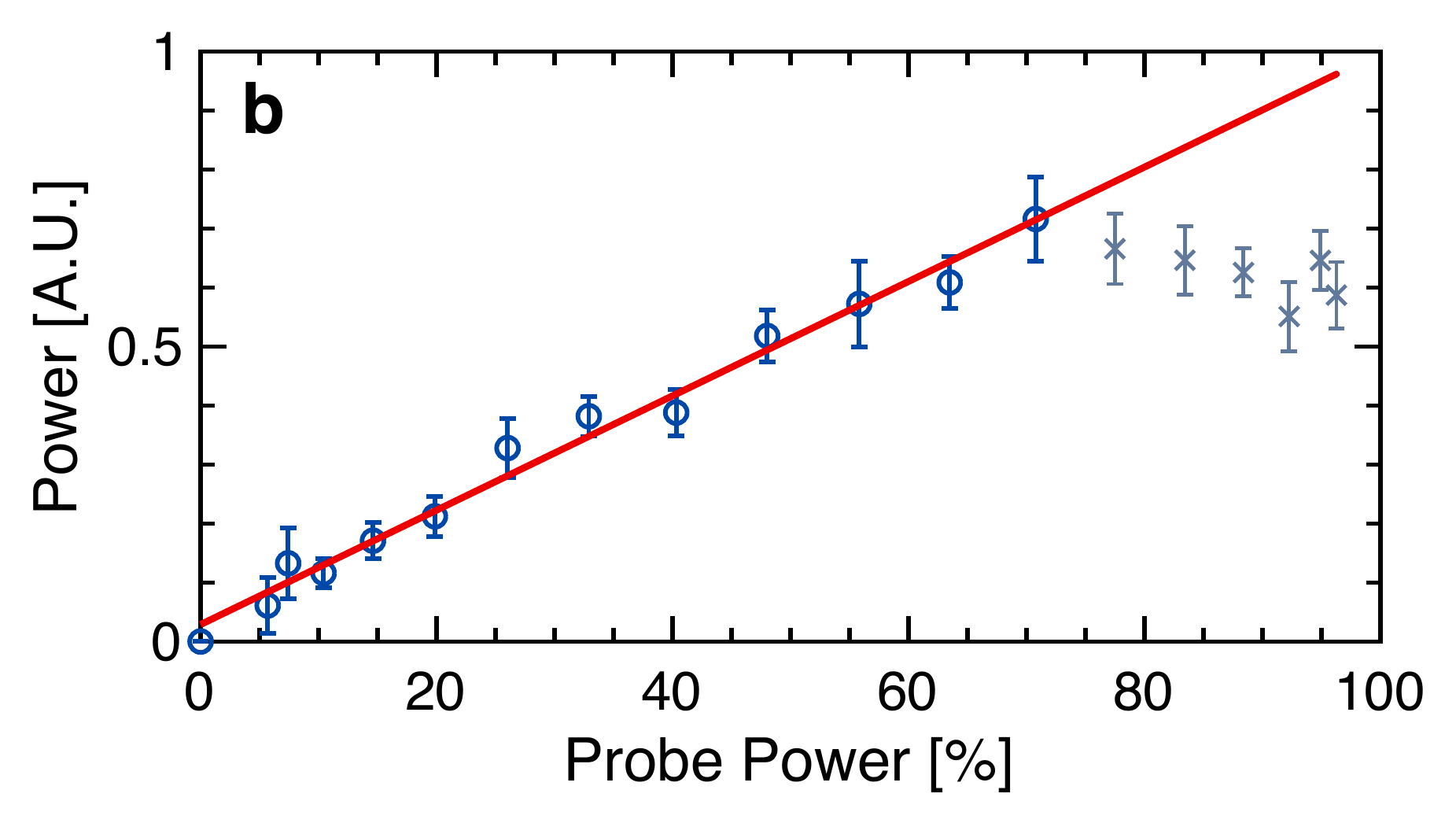} \label{test b}}
\caption{Experimental checks. {\bf a}: Comparison with theory. Measured smoothed peaks (circles) of the stimulated negative Hawking radiation (NHR,  see {\it e.g.}\ Fig.~\ref{results}c) for various probe wavelengths versus theory (curve, from Fig.~\ref{dopplercurve}). The error bars indicate our spectral resolution. The outlier is due to a complicated peak structure. {\bf b:} Linearity of stimulated Hawking radiation. Power of the UV signal (Fig.~\ref{results}c) for \SI{1600}{\nano\meter} probe wavelength as a function of probe power (circles and crosses) versus a linear fit (for the circles). The maximal probe power reaches $\approx 2\%$ of the peak intensity of the pump. The figure shows that the power of the generated radiation is linear in the probe power for low intensities, whereas for higher power it saturates.}
\label{dispersion}
\end{figure}

To test this hypothesis, we vary the probe wavelength and compare (Fig.~\ref{test a}) each UV peak, after subtraction of the pump contribution, with the theoretical prediction (Fig.~\ref{dopplercurve}) based on the Doppler formula (\ref{doppler}) with the effective refractive index $n(\omega)$ that depends on both the material and the microstructure of the fiber \cite{Agrawal}. We obtain $n(\omega)$ from measurements of the group index in the IR and Visible interpolated to the material refractive index in the UV \cite{Details} checked and fine--tuned with our measurement of the previously known negative resonant radiation \cite{Faccio2}. The group velocity $u$ of the pump was fitted and corresponds to a carrier wavelength of \SI{818.9}{\nano\meter}, consistent with measurements \cite{Details}. The good agreement with the prediction (Fig.~\ref{dopplercurve}) we take as evidence for the correct interpretation of the new UV peak as stimulated Hawking radiation. Further supporting evidence is given by numerical calculations of the pump--probe interaction in the negative $\omega'$ range \cite{Details}.

Additionally, we have also varied the probe power while keeping everything else constant. Figure~\ref{test b} shows that the power of the UV peak due to the probe is linear in the probe intensity for low probe power until it saturates for a probe power of $\approx 1.5\%$ of the pump peak power. The linearity is another important feature of a stimulated effect, while the saturation indicates a regime known from numerical simulations \cite{Shalva} where the probe is able to influence the pump with relatively low power --- where probe and pump switch sides.

We have thus strong reasons for the correct interpretation of the observed UV peak (Fig.~\ref{results c}) as stimulated Hawking radiation: the agreement with the Doppler formula (\ref{doppler}) for negative frequencies in the co--moving frame with the measured and calculated refractive--index data \cite{Details}, supplemented with numerical simulations \cite{Details}, and the linearity of the stimulated signal (Fig.~\ref{test b}) for low probe power. The measurements show empirically where the spectrum of the stimulated Hawking radiation lies, and hence also where the spontaneous Hawking radiation is expected. However, our pump--probe technique does not allow us to make precise measurements of the Hawking spectrum, as the probe spectrum is too wide. We do not expect \cite{David} a Planck spectrum there, as we are in a Hawking regime of strong dispersion \cite{Criterion}.  We also found \cite{Details} that the UV part of the stimulated Hawking radiation consists of multiple modes. 

We can estimate how many Hawking quanta are spontaneously produced in the mode we detect with our current apparatus. For a probe intensity of \SI{1}{\kilo\watt} we have $2\times 10^7$ photons \cite{Estimation} stimulating $41,000$ additional counts per second between \SI{229.7}{\nano\meter} and \SI{231.5}{\nano\meter} (Fig.~\ref{results}c), which corresponds to $2\times 10^{-3}$ spontaneous UV Hawking partners per second (detected with our current efficiency of about $10^{-2}$). In the IR the fiber is single--mode, concentrating all light into a guided wave, and there numerics \cite{Details} indicate a $10^4$ times higher Hawking rate.

Our measurements prove that the optical analogue of the event horizon \cite{Philbin} does indeed describe our observations, despite other effects present in nonlinear fiber optics \cite{Agrawal} such as third--harmonic generation and the Raman effect. Third harmonics \cite{Agrawal} are produced, because the nonlinear polarization is proportional to the cube of the instant electric field. This gives two contributions: while one appears as the refractive--index perturbation $\delta n$ we use for generating Hawking radiation, the other oscillates at trice the carrier frequency of the pulse and generates third harmonics. We have seen the wide, unstructured range of non--resonant third harmonics over the third of the wavelength range of the pulse, but both the negative--frequency peak of the pump and the stimulated Hawking radiation of the probe lie at the tail of this range and are clearly distinguishable (Fig.~\ref{results}c). 

The Raman effect  \cite{Agrawal} de--accellerates the pump pulse, which makes the pulse velocity intensity--dependent, and hence also the horizon. However, most of the stimulated Hawking radiation is generated during the first \SI{1}{\milli\meter} of propagation in the fiber. There the pump pulse, of soliton number $N=2.2$ \cite{Agrawal}, gets compressed to almost an optical cycle, before it splits into two solitons \cite{Details}. During this short propagation distance the Raman effect is small, shifting the carrier wavelength from \SI{800}{\nano\meter} to about \SI{820}{\nano\meter} \cite{Details}. In the blue and red shifting of the probe (Figs.~\ref{results}a and \ref{results}b) the Raman effect is more significant where the horizon wavelength (\SI{1551}{\nano\meter}) is substantially shifted compared with the prediction (\SI{1613}{\nano\meter}) based on the refractive--index data (Fig.~\ref{dispersion}b) \cite{Details}. 

It is quite remarkable that both the violent pulse dynamics and the other effects of nonlinear fiber optics \cite{Agrawal} are not affecting the essential physics of the optical event horizon \cite{Philbin,Book}, which is a prerequisite for the next milestone: the optical observation of quantum Hawking radiation. In addition, this robustness and the demonstration of probe--controlled extreme frequency conversions --- between positive and negative frequencies  --- seem to appear as important insights on their own.

{\it Acknowledgements.---} We thank
Shalva Amiranashvili,
Uwe Bandelow,
Ora Bitton,
Itay Griniasty,
Theodor W. H\"{a}nsch, 
Sunil Kumar,
Michael Kr\"{u}ger,
Rafael Probst, 
Arno Rauschenbeutel,
and
Thomas Udem
for help and helpful discussions. 
Our work was supported by the European Research Council, the Israel Science Foundation, a research grant from Mr. and Mrs. Louis Rosenmayer and from Mr. and Mrs. James Nathan, and the Murray B. Koffler Professorial Chair. J. D. and Y. R. contributed equally to this work.


\begin{thebibliography}{99}
\bibitem{Oom}
The maximal mass of a black hole made in the gravitational collapse of stars is bounded by the Chandrasekhar limit of $\approx 1.39$ solar masses, which implies that the Hawking temperature \cite{Hawking} lies below $5\times 10^{-8}\mathrm{K}$. This is significantly below the temperature of the cosmic microwave background and its fluctuations.
\bibitem{Note}
Note that this frequency shift has not been taken into account in the Bogoliubov transformation with constant coefficients, Eq.~(\ref{bogo}), for simplicity. It could be included in each positive outgoing mode $\hat{b}_+$ by a prefactor oscillating with this shift, but the shift is small compared with the gap between positive and negative frequencies.
\bibitem{Criterion}
The pump pulse produces a perturbation $\delta n \approx 10^{-3}$, whereas the group and phase velocity differ by $\approx 10^{-2}$, which indicates a regime of strong dispersion where the Hawking spectrum is no longer Planckian \cite{David}.
\bibitem{Estimation}
For a relative group--velocity difference matching the perturbation $\delta n$ \cite{Criterion} about \SI{3}{\femto\second} of probe light interacts with the pump along \SI{1}{\milli\meter}, which for \SI{1}{\kilo\watt} contains $2\times 10^7$ photons of \SI{1.5}{\micro\meter} wavelength.
\end{thebibliography}
\end{document}